\documentclass[a4paper,twocolumn,american]{revtex4}
\usepackage[T1]{fontenc}
\usepackage[latin1]{inputenc}
\usepackage{graphicx}

\makeatletter

\providecommand{\tabularnewline}{\\}

\usepackage{times}

\usepackage{babel}
\makeatother
\begin{document}

\title{Anomalous transport in Charney-Hasegawa-Mima flows}

\author{Xavier Leoncini}

\affiliation{{\small PIIM, Université de Provence, CNRS, Centre Universitaire
de Saint Jérôme, F-13397 Marseilles, France}}

\email{Xavier.Leoncini@up.univ-mrs.fr}

\author{Olivier Agullo}

\affiliation{{\small PIIM, Université de Provence, CNRS, Centre Universitaire
de Saint Jérôme, F-13397 Marseilles, France}}

\email{agullo@up.univ-mrs.fr}

\author{Sadruddin Benkadda}

\affiliation{{\small PIIM,  Université de Provence, CNRS, Centre Universitaire
de Saint Jérôme, F-13397 Marseilles, France}}

\email{benkadda@up.univ-mrs.fr}

\author{George M. Zaslavsky}

\affiliation{{\small Courant Institute of Mathematical Sciences, New York University,
251 Mercer St., New York, NY 10012, USA }}

\affiliation{{\small Department of Physics, New York University, 2-4 Washington
Place, New York, NY 10003, USA }}

\email{zaslav@cims.nyu.edu}

\begin{abstract}
{\footnotesize Transport properties of particles evolving in a system
governed by the Charney-Hasegawa-Mima equation are investigated. Transport
is found to be anomalous with a non linear evolution of the second
moments with time. The origin of this anomaly is traced back to the
presence of chaotic jets within the flow. All characteristic transport
exponents have a similar value around $\mu=1.75$, which is also the
one found for simple point vortex flows in the literature, indicating
some kind of universality. Moreover the law $\gamma=\mu+1$ linking
the trapping time exponent within jets to the transport exponent is
confirmed and an accumulation towards zero of the spectrum of finite
time Lyapunov exponent is observed. The localization of a jet is performed,
and its structure is analyzed. It is clearly shown that despite a
regular coarse grained picture of the jet, motion within the jet appears
as chaotic but chaos is bounded on successive small scales.}{\footnotesize \par}
\end{abstract}


\maketitle

\section{Introduction}

Understanding transport in turbulent magnetized plasma is a task of
fundamental importance. In these plasma, transport problems are often
related to confinement, which is one of the last standing issues confronting
the realization of magnetically confined controlled fusion devices.
As the literature evolves, there has been more and more evidence showing
that the transport properties can be anomalous, in the sense that
transport may not be correctly described by Gaussian kinetics, but
by what one now calls {}``strange kinetics'' \cite{Schlesinger93}.
In these regards, transport phenomena in turbulent magnetized plasma
is part of the ever growing number of physical systems displaying
anomalous properties\cite{Carreras03,Annibaldi00,Castillo2004,Grandgirard2000}.
As of today the full understanding of these phenomena are far from
being complete and in many regards a full blown theory able to capture
and describe correctly these {}``strange kinetics'' has not yet
surfaced. There seems nevertheless to be a common agreement to link
these phenomena to Levy-type processes and their generalizations.
Moreover the use of fractional derivatives in Fokker-Plank-Kolmogorov
type equations captures qualitatively some of the transport properties
and is thus a good step towards a proper description of anomalous
transport \cite{Zaslavsky2002}.

The link between the Hamiltonian dynamics and the kinetics at origin
of anomalous transport properties are relatively well understood when
dealing with low dimensional systems such as a time periodic flow
which belong to the class of $1-1/2$ degree of freedom Hamiltonians.
The dynamics in these systems is not ergodic: a well-defined stochastic
sea, with chaotic dynamics, filled with various islands of quasi-periodic
dynamics compose the phase space. The anomalous properties and their
multi-fractal nature are then linked to the existence of islands within
the stochastic sea and the phenomenon of stickiness observed around
them \cite{Kuznetsov2000,LKZ01}. However when dealing with more complex
systems the loss of time-periodicity complicates the picture. For
instance in geophysical flows or two-dimensional plasma turbulence,
the islands which were static and well localized in phase space, are
replaced by {}``coherent structures'', which have a life of their
own. Hence, tackling the origin of anomalous transports from the chaotic
dynamics of individual tracers becomes more subtle. Recently, the
existence of a hidden order for the tracers which exhibits their possibility
to travel in each other's vicinity for relatively large times was
exhibited \cite{LZ02}. This order is related to the presence within
the system of chaotic jets \cite{Afanasiev91,LZ03}. These chaotic
jets can be understood as moving clusters of particles within a specific
domain for which the motion appears as almost regular from a coarse
grained perspective. Typically, the chaotic motion of the tracers
is confined within the characteristic scale of a given jet, within
which nearby tracers are trapped. From another point of view, looking
for chaotic jets can be understood as a particular case of measurements
of space-time complexity\cite{Afraimovich03}.

The purpose of this paper is to study transport properties and to
look for chaotic jets in a model of two-dimensional turbulence which
applies either in the plasma context where it is known as the Hasegawa-Mima
equation \cite{Hasegawa1981} or in the geophysical one where one
speaks about the Charney equation (see for instance \cite{Pedlosky_book}).
In this setting Annibaldi et al have already shown strong evidence
of anomalous transport of passive particles \cite{Annibaldi00,Annibaldi02},
hence the search for chaotic jet is expected to give some clues on
the origin of anomalous transport in these systems, and for instance
to identify the structures responsible for such transport. Indeed
in \cite{LZ02}, it was clearly shown that trapping within chaotic
jets resulted in anomalous transport. Moreover it appeared that jets
were localized around the coherent structures of systems, namely the
vortex cores and that the structure of the jets itself was a hierarchy
of jets within jets, reminiscent of the multi-fractal nature of transport
observed in $1-1/2$ degree of freedom Hamiltonians systems \cite{LKZ01}.
Hence, the goal of this paper may be seen as two-fold. First we want
to understand the origin of anomalous transport in a model of two-dimensional
plasma turbulence. And second, since we are using chaotic jets to
track this origin, we are at the same time testing the existence of
these jets in a more complex setting than the system of point vortices
used in \cite{LZ02}. Once the presence of jets is confirmed we may
be able to speculate that the different anomalous transport behavior
portrayed in the nonexhaustive following references \cite{Nau99,Carreras03,Dickman04,Dupont98,Sukhatme04,Beyer2001}
may all find their origin in the presence of long lived chaotic jets
in the considered systems. 

The paper is organized as follows. In Sec. \ref{sec:Basic-definitions}
the basic definitions are introduced. A brief introduction of the
Charney-Hasegawa-Mima equation is given. Then the dynamics of test
particles is given and the Lagrangian approach of transport in this
setting is discussed. In Sec. \ref{sec:Transport-properties} the
dynamical evolution of the field and of passive test particles, as
well as transport properties are computed numerically. First the numerical
setting is discussed. Then three different regimes corresponding to
different choices of the parameters for the field evolution are considered
cases. In Sec. \ref{sec:Jets} we track the origin of anomalous transport
in the three considered. First we recall the definition of a chaotic
jet and present the numerical method used to detect these jets. We
then present the statistical results related to trapping time within
jets and analyze the origin of anomalous transport by localizing the
jets and by describing their structure. Finally we conclude in Sec.
\ref{sec:Conclusion}.

\section{Basic definitions\label{sec:Basic-definitions}}

\subsection{The Charney-Hasegawa-Mima equation}

The Charney-Hasegawa-Mima equation can be written in the following
form,

\begin{eqnarray}
\partial_{t}\Omega & +[\Omega,\Phi] & =0\label{eq:H-M}\\
\Omega & = & \Phi-\lambda\Delta\Phi+gx\:,\label{eq:Gen_Vorticity}\end{eqnarray}
where $[\cdot,\cdot]$ corresponds to the Poisson operator, $\Omega$
is a generalized vorticity given by Eq.(\ref{eq:Gen_Vorticity}),
$\Phi$ is a potential and $\lambda$ and $g$ are parameters.

Typically Eqs. (\ref{eq:H-M}) and (\ref{eq:Gen_Vorticity}) can either
describe the evolution of an anisotropic plasma, and are then referred
to as the Hasegawa-Mima equation, or the evolution of geostrophic
flows in which context they are known as the Charney equation. This
formal identity has an advantage as the results obtained in this paper
for transport properties should apply in either context.

It is however important to be able to put the results back in a physical
context. With this in mind we provide in the next section a short
derivation of Eqs (\ref{eq:H-M}) and (\ref{eq:Gen_Vorticity}) in
both the anisotropic plasma configuration and the geostrophic approximation.

\subsection{Wave-vortex paradigm equation for two-dimensional flows.}

\subsubsection{The Hasegawa-Mima equation}

Let us start by considering a magnetized anisotropic plasma, \emph{e.g}
a plasma with a uniform magnetic field along a direction, $\mathbf{B}=B\mathbf{z}$.
We shall also assume that electron response to the turbulent fluctuations
of the electric potential $\phi$ is adiabatic, $n_{e}(x,y,t)=n_{0}(x,y)e^{e\phi(x,y,t)/T_{e}}$,
where $n_{e}$ is the electron density, $n_{0}$ is the equilibrium
plasma density, and $T_{e}$ the electron temperature. In the anisotropic
plasma let the cyclotron frequency be $\omega_{c}=eB/m_{i}$, and
let the hybrid sound speed be $c_{s}=\sqrt{T_{e}/m_{i}}$, where $m_{i}$
is the ion's mass.

We now consider the motion of cold ions (the ion temperature is assumed
zero, $T_{i}=0$) on characteristic time and length scales much larger
respectively than $1/\omega_{c}$ and the Debye length $\lambda_{D}$.
In this situation the plasma is quasi-neutral $n{}_{i}\sim n_{e}$,
and a linear combination of ion continuity equation and momentum equations
gives (see \cite{agullo2003} for details)

\begin{equation}
-\nabla.\mathbf{v}_{\bot}=\frac{D}{Dt}\left(\frac{e\phi}{T_{e}}\right)+\mathbf{v}_{\bot}.\nabla\ln n_{0}=\frac{1}{\omega+\omega_{c}}\frac{D}{Dt}(\omega+\omega_{c})\;,\label{eq:div(v) has-mima}\end{equation}
where $\mathbf{v}_{\bot}$ is the speed in the plane perpendicular
to the magnetic field and $\mathbf{\omega}=\nabla\times\mathbf{v}_{\bot}$
is the vorticity of the 2D flow. The slow motion of the ions necessarily
implies that $\mathbf{v}_{\bot}\sim\mathbf{v}_{D}=B_{0}^{-2}\mathbf{B_{0}}\times\nabla\phi$
and, using the notation $\varphi\equiv e\phi/T_{e}$, subtraction
of the right hand side terms in (\ref{eq:div(v) has-mima}) readily
gives \begin{equation}
\frac{D}{Dt}(\triangle\varphi-\varphi/l_{L}^{2})-\mathbf{v}_{D}.\nabla\ln n_{0}=0\;,\label{eq:has_mima n_0(x,y)}\end{equation}
where $l_{L}=c_{s}/\omega_{c}$ is the hybrid Larmor radius.

When the plasma is homogeneous ($n_{0}$=const.), the turbulent dynamics
of (\ref{eq:has_mima n_0(x,y)}) is characterized by the absence of
waves and the formation of large vortices. Moreover, in the $l_{L}\rightarrow+\infty$
limit, Eq. (\ref{eq:has_mima n_0(x,y)}) becomes formally equivalent
to the 2D momentum Euler equation ($\varphi$ being the stream function
in the Euler case). 

In plasma devices, density is larger in the core than at the boundaries.
It this setting we can often write $n_{0}(x,y)\sim n_{0}e^{-x/L_{n}}$,
where $L_{n}$ is a characteristic density gradient length. Using
this last approximation we finally obtain the Hasegawa-Mima equation:\begin{equation}
\frac{D}{Dt}(\triangle\varphi-\varphi/l_{L}^{2})-g\frac{\partial\varphi}{\partial y}=0\;,\label{eq:has_mima}\end{equation}
where $g=\omega_{c}/L_{n}$.

The inhomogeneous character of the equilibrium density profile in
Eq.(\ref{eq:has_mima}) implies the existence of so called drift-waves
in the flow. In particular, the waves deform and interact with the
self-organized vortex-like or multipole-like structures, and play
a key role in their interactions. For instance the collision of two
dipoles can lead to two monopoles plus one dipole plus some radiation
\cite{Fontan1995}.

\subsubsection{The Charney Equation}

Let us now explain the physical nature of the intrinsic length in
atmospheric motions. In the context of shallow water approximation
where the density $\rho$ is uniform, the modeling of the evolution
of the atmospheric winds leads to a dynamic equation similar to (\ref{eq:div(v) has-mima})
or (\ref{eq:has_mima}). Indeed, the atmosphere is characterized by
small Ekman and Rossby numbers \cite{Pedlosky_book}, so that the
motion of thin incompressible fluid rotating layer (thin with respect
to the characteristic scale $L_{\bot}$ of the horizontal motion)
is such that friction forces are very weak compared to the inertial
Coriolis force, second and that pressure forces are balanced by gravity
in the vertical direction $\mathbf{z}$ (hydrostatic approximation).
As a consequence the horizontal pressure gradient is independent of
the vertical component $z$. It is also proportional to the layer
width $h(x,y,t)$ (because of the free boundary condition $p(x,y,h)=p_{0}=$
const.). It is therefore reasonable to assume that the horizontal
velocity field $\mathbf{v}_{\bot}$ is also $z$-independent: $\mathbf{v}=\mathbf{v}_{\bot}(x,y,t)+w(x,y,z,t)\mathbf{z}$. 

If we assume that the rigid earth surface is locally flat, i.e $w(z=0)=0$,
integration of the incompressibility condition leads to $w=-z\nabla.\mathbf{v}_{\bot}$and,
in particular, $dh/dt=-h\nabla.\mathbf{v}_{\bot}$. Moreover, since
the forces acting on the fluid are the Coriolis and pressure forces,
$\rho f\mathbf{v}_{\bot}\times\mathbf{z}$ and $-\rho\nabla p$, it
is straightforward to deduce from the momentum equation the exact
identity $d(\omega+f)/dt=(\omega+f)\nabla.\mathbf{v}_{\bot}$. It
follows that \begin{equation}
-\nabla.\mathbf{v}_{\bot}=\frac{D}{Dt}\left(\frac{h}{H_{0}}\right)=\frac{1}{\omega+f}\frac{D}{Dt}(\omega+f)\;,\label{eq:div(v) Charney}\end{equation}
where we put $h\equiv h+H_{0}$, $H_{0}$ being the mean width of
the layer and $h$ the fluctuations around the mean ($h\ll H_{0}$
by hypothesis). The Coriolis term $f\mathbf{z}$ is the local component
of the planetary vorticity $\omega_{planet}$ in the vertical direction,
the northward local component $f_{n}\mathbf{y}$ being negligible.
Indeed, the Coriolis force in the horizontal direction is $(-2\rho\omega_{planet}\times\mathbf{v})_{\bot}=\rho\mathbf{v}_{\bot}\times f\mathbf{z}+\rho w\mathbf{z}\times f_{n}\mathbf{y}\sim\rho\mathbf{v}_{\bot}\times f\mathbf{z}$
because, in the shallow water approximation, $w/v_{\bot}=O(H_{0}/L_{\bot})\ll1$
and far from equator latitudes, $f_{n}\sim f$.

It is clear from equations (\ref{eq:div(v) has-mima}) and (\ref{eq:div(v) Charney})
that in the special limits where the plasma is homogeneous ($n_{0}=$const.)
and the Coriolis parameter $f=f_{0}$ is taken constant, both physical
systems are formally equivalent. We just have to identify $e\phi/T_{e}$with
$h/H_{0}$ and choose as length and time units $(l_{L},1/\omega_{c})$
or $(l_{R},1/f_{0})$, where $l_{R}=\sqrt{gH_{0}}/f_{0}$ is the so-called
Rossby length. But the similarity holds even for the case of a layer
with a slow dynamics ($\omega\ll f$) by taking into account the slow
variation of the Coriolis parameter in the northward $y$-direction:
$f=f_{0}+\beta y$ with $\delta f/f_{0}\ll1$ ($\beta$-plane approximation).
In fact, since the Rossby number is small, we can also make the approximation
$\mathbf{v}_{\bot}\sim\mathbf{v}_{D}=gf^{-1}\mathbf{z}\times\nabla h$.
And by subtracting the right hand side (r.h.s) terms of (\ref{eq:div(v) Charney}),
we obtain the so-called Charney equation\begin{equation}
\frac{D}{Dt}(\triangle h-h/l_{R}^{2})-\beta\frac{\partial h}{\partial y}=0\;.\label{eq:Charney}\end{equation}
 It is clear that (\ref{eq:has_mima}) and (\ref{eq:Charney}) are
formally equivalent. However, in the atmospheric case, waves are not
generated by a density inhomogeneity but by the earth rotation. As
mentioned earlier, both equations (\ref{eq:has_mima}) and (\ref{eq:Charney})
can be written in the compact form given by (\ref{eq:H-M}) where
Eq.~(\ref{eq:Gen_Vorticity}) actually writes $\Omega=\varphi-l_{L}^{2}\Delta\varphi-x/L_{n}$
(the transformations $\varphi\hookrightarrow\textrm{h}$, $g\hookrightarrow\beta$,
and $l_{L}\hookrightarrow l_{R}$ lead to the Charney equation).

\subsection{Advection equation}

For an incompressible fluid, the evolution of a passive particle is
given by the advection equation\begin{equation}
\dot{z}=v(z,t)\label{gen.adv}\end{equation}
 where $z(t)$ represent the position of the tracer at time $t$ in
the complex plane, and $v(z,t)$ is the velocity field. An important
feature of this evolution of passive particles is that the evolution
equation given by Eq.~(\ref{gen.adv}) can be rewritten as:\begin{equation}
\dot{z}=-i\frac{\partial\Phi}{\partial\bar{z}}\:,\hspace{10mm}\dot{\bar{z}}=i\frac{\partial\Phi}{\partial z}\:,\label{eq:Tracers_Hamiltonian}\end{equation}
where the potential $\Phi$ acts as a time dependent Hamiltonian,
and $\bar{}$ denotes the complex conjugate. This Hamiltonian structure
is fundamental as it imposes some constraints on the dynamics of passive
tracers, which should be taken into account when carrying a numerical
simulation.

\subsection{Numerical settings}

\subsubsection{Charney-Hasegawa-Mima}

Simulations of the Charney-Hasegawa-Mima equation are performed for
different initial conditions and choices of parameters. The choices
are made to be consistent with the literature, namely the conditions
chosen in \cite{Annibaldi00,Annibaldi02}. The simulations are performed
within a square box of size $L=20$ and periodic boundary conditions
using a pseudo-spectral code. In order to compute the evolution of
passive tracers accurately we settled for a somewhat low resolution
mesh of $128^{2}$. Fourier transforms are computed using a fast Fourier
algorithm. For the time evolution, we chose a $4$th-order Runge-Kunta
integration scheme with typical time step $\delta t=0.05$.

In order to avoid numerical instability, as well as a trivial asymptotic
behavior Eq.~(\ref{eq:H-M}) could not be kept as it is and a dissipation
term $D$ as well as a forcing term $F$ were added:\begin{equation}
\partial_{t}\Omega+[\Omega,\Phi]=D+F\:.\label{eq:Mima_forced}\end{equation}
One may also argue that the dissipation term may be relevant to describe
some physical phenomena.

For the dissipation we used a hyper-viscous term $D=\nu(\nabla\Omega)^{4}$.
For the forcing we used a term, whose Fourier transform is:\begin{equation}
\hat{F}(k_{x},k_{y})=F_{0}\frac{1}{\sqrt{\nu\sum_{k}k^{4}}}e^{i\varphi(k_{x},k_{y})}\:,\label{eq:forcing_term}\end{equation}
where for any $(k_{x},k_{y})$, $\varphi(k_{x},k_{y})$ is a random
phase uniformly distributed on the circle; $k$ runs over an intervall
centered on $k_{0}$ with a range $\delta k$. One can set also a
time dependency by regularly updating the realization of the random
phases. 

The initial condition is given by the following choice for the $\Omega'=\Phi-\lambda\Delta\Phi$
field:\begin{equation}
\Omega'=A_{0}\sum\frac{1}{\sqrt{m^{2}+n^{2}}}\sin\left(\frac{2\pi m}{L}\left(x-\frac{L}{2}\right)\right)\cos\left(\frac{2\pi n}{L}y+\varphi_{i,j}\right)\:,\label{eq:Init_cond}\end{equation}
where $\varphi_{i,j}$ is a random phase uniformly distributed on
the circle.

\subsubsection{Passive tracers}

It is important to take special care of the way the dynamics of the
tracers are computed to characterize the possible anomalous properties
of transport. If they exist, such anomalies should find their origin
in the existence of {}``memory effects'', namely long time correlations,
since the accessible range of velocities is finite. In this perspective,
any source of randomness leading to memory loss due to the numerical
scheme may then induce a spurious effective diffusive behavior. Moreover
the Hamiltonian nature of the tracers dynamics imposes necessarily
the choice of a simplectic integrator.

We thus chose the sixth-order implicit Gauss-Legendre simplectic scheme
to compute the trajectories \cite{McLachlan92}, as this integration
scheme was successfully used in systems of point vortices \cite{Kuznetsov98,Laforgia01,LKZ01,LZ02}.
However in order to avoid a possible source of noise, we had to compute
the velocities of particles {}``exactly'', meaning that we performed
an exact back-Fourier transform of the modes describing the evolution
of the field. This constraint is numerically expensive, and explains
our choice of a relatively low resolution of $128^{2}$ for the evolution
of the field. In this setting the evolution of passive tracers may
be understood as describing the advection of particles in a flow field
generated by $128^{2}$ modes interacting through Eq.~(\ref{eq:H-M}).

\subsection{Field settings}

As to the choice of parameters and initial conditions in Eq.~(\ref{eq:H-M})
we considered three different cases, with quite different values of
the parameters. As mentioned earlier we chose as a starting base similar
initial conditions, type of forcing, and size of the system as in
\cite{Annibaldi00}. For the three considered cases, for given initial
conditions, we let the system evolve until it becomes sufficiently
stationary for the time span considered during simulation, $\tau_{final}=10^{4}$.
Stationarity is considered as being reached by monitoring the evolution
of the energy and enstrophy of the system (see Table~\ref{table:Field_table}).
When forcing is present (case 2 and 3), the random phases of the forcing
are updated all at once every $\delta\tau=2$. The values assigned
to $A_{0}$ are respectively for the three cases: $1.1$, $0.001$,
$0.5$. For the first case, as there was no forcing, we tried a decent
amplitude, for the second case with strong forcing we let the forcing
drive the system and for the last case, we tried something intermediate.
The motivation for these choice was to get three different stationary
fields, as the focus of this paper is actually more the transport
of test particles rather than the actual dynamics of the field.

In order to visualize the field we chose to use levels of the function
$-\Delta\Phi$. The three different considered cases are represented
in figures \ref{cap:Case1}, \ref{cap: Case 2} and \ref{cap: Case 3}.
\begin{table}[htbp]

\caption{\label{table:Field_table} Density of Energy and Enstrophy for the
three considered cases}

\begin{center}\begin{tabular}{c|c|c|}
&
Energy&
Enstrophy\tabularnewline
\hline
Smooth Field&
$0.64\pm0.01$&
$0.8\pm0.05$\tabularnewline
\hline 
Forced Field&
$2.62\pm0.005$&
$10.81\pm0.03$\tabularnewline
\hline 
Anisotropic Field&
$0.28\pm0.005$&
$0.18\pm0.01$\tabularnewline
\hline
\end{tabular}\end{center}
\end{table}

\subsubsection{Smooth Field}

\begin{figure}
\includegraphics[%
  width=7cm,
  keepaspectratio]{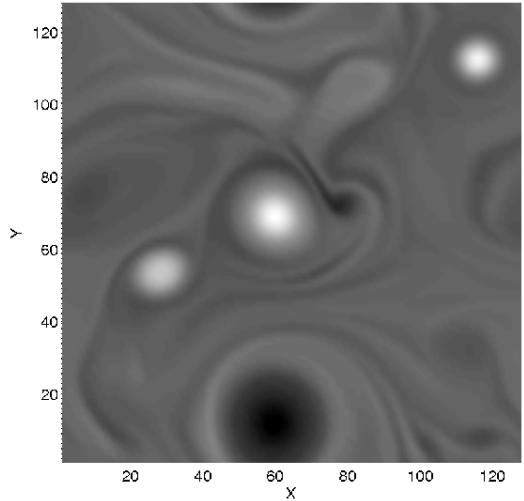}

\caption{Visualization of the field $-\Delta\Phi$ for the choice of parameters
$\lambda=1$, $g=0.1$, $\nu=7\;10^{-6}$, $L=20$, $N=128^{2}$ with
no forcing. The field is {}``smooth'' and appears as being isotropic.
A few vortices are present. The gray coloring scheme scales from -4.5
(black) to 4.5 (white)\label{cap:Case1}}
\end{figure}
To obtain the {}``smooth'' field depicted in Fig. \ref{cap:Case1},
we carried out simulations with no forcing $(F_{0}=0)$ and a low
dissipation. The parameters for this run were $\lambda=1$, $g=0.1$,
$\nu=7\;10^{-6}$. Due to this low dissipation the energy may be considered
as constant for the length of the simulation. We notice that a few
distinct vortices are present. During the evolution a merger between
two vortices occurred. There is also an average drift in the y-direction.

\subsubsection{Forced Field}

\begin{figure}
\includegraphics[%
  width=7cm]{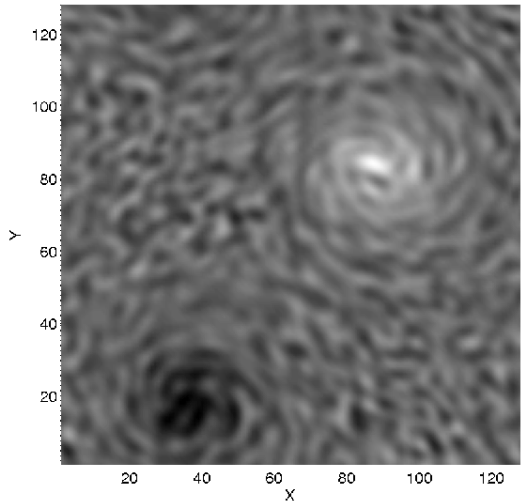}

\caption{Visualization of the field $-\Delta\Phi$ for the choice of parameters
$\lambda=4$, $g=0.1$, $\nu=5\;10^{-5}$, $F_{0}=4$, $k_{0}=6\pm2$
, $\Delta t=2$. Two big perturbed vortices are present. The gray
coloring scheme scales from -1.4 (black) to 1.4 (white)\label{cap: Case 2}}
\end{figure}

To obtain the {}``forced'' field depicted in Fig. \ref{cap: Case 2},
we carried out simulations with a strong forcing and dissipation,
the parameters for this run were $\lambda=4$, $g=0.1$, $\nu=5\;10^{-5}$,
$F_{0}=4$, $k_{0}=6\pm2$,%
 the phases for the random forcing were updated every $\Delta\tau=2$
time units. The value of $k_{0}$ corresponds to physical scales of
$\delta x\sim\frac{2\pi}{k_{0}}\approx1$. With this choice of parameters
the system consists of two perturbed vortices. An average drift in
the y-direction is also noticeable.

\subsubsection{Anisotropic Field}

\begin{figure}
\includegraphics[%
  width=7cm]{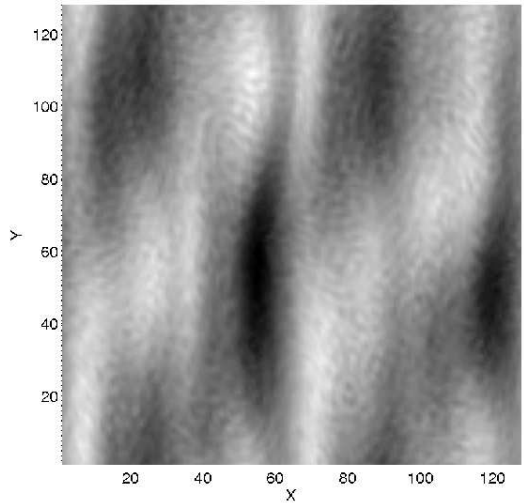}

\caption{Visualization of the field $-\Delta\Phi$ for the choice of parameters
$\lambda=0.125$, $g=2$, $\nu=7.5\;10^{-6}$, $F_{0}=1.5$, $k_{0}=12\pm2$,
$\Delta t=2$. Elongated structures in the y-direction are present,
marking a strong anisotropy. The gray coloring scheme scales from
-1.6 (black) to 1.6 (white)\label{cap: Case 3}}
\end{figure}

To obtain the {}``anisotropic '' field depicted in Fig. \ref{cap: Case 3},
we carried out simulations with some forcing and a high value for
$g$. The parameters for this run were $\lambda=0.125$, $g=2$, $\nu=7.5\;10^{-6}$,
$F_{0}=1.5$, $k_{0}=12\pm2$. The value of $k_{0}$ corresponds to
physical scales of $\delta x\sim\frac{2\pi}{k_{0}}\approx0.5$. The
phases for the random forcing are updated every $\Delta\tau=2$ time
units. In this settings elongated structures as well as a strong drift
in the y-direction is present. %

\section{Transport properties\label{sec:Transport-properties}}

\subsection{Definitions}

Unfortunately the deterministic description of the motion of a passive
particle in a chaotic region is impossible. Local instabilities produce
exponential divergence of trajectories. Thus even an idealized numerical
experiment is non-deterministic, as round-off errors are creeping
slowly but steadily from the smallest to the observable scale. The
long-time behavior of tracer trajectories is then necessarily studied
by using a probabilistic approach. In the absence of long-term correlations,
the kinetic description, which uses the Fokker-Plank-Kolmogorov equation,
leads to Gaussian statistics. Yet if a phenomenon with associated
long time correlations occurs, profound changes in the kinetics can
be induced. These memory effects sometimes result in the modification
of the diffusion coefficient in the FPK equation \cite{Chirikov79,Rechester80}.
But often their influence is more profound \cite{zaslavsky93,DcN98,Castiglione99,Kuznetsov2000,LKZ01,LZ02},
and leads to non-Gaussian statistics, and for instance to a non-diffusive
behavior of the particle displacement variance: \begin{equation}
\langle(s-\langle s\rangle)^{2}\rangle\sim t^{\mu}\:,\label{anom}\end{equation}
where $\langle\cdots\rangle$ stands for ensemble averaging. Within
this probabilistic approach, the main observables in order to characterize
transport properties will be moments of the distributions:\begin{equation}
M_{q}(t)=\langle|s(t)-\langle s(t)\rangle|^{q}\rangle\:,\label{momentsdefi}\end{equation}
where $q$ denotes the moment order. The finiteness of velocity and
of time in our simulations implies that all moments are finite and
that a power law behavior is expected\begin{equation}
M_{q}\sim D_{q}t^{\mu(q)}\:.\label{momentexpectation}\end{equation}
with, generally, $\mu(q)\ne q/2$ as would be expected for normal
diffusion. The nonlinear dependence of $\mu(q)$ is a signature of
the multifractality of the transport, while its linear dependence
reflects a fractal situation \cite{Castiglione99,Andersen2000,Ferrari01}.
In the fractal situation all of the moments can be described by a
single self-similar exponent $\nu$ \begin{equation}
\mu(q)=\nu\cdot q\:,\label{eq:weak}\end{equation}
 whereas the case when $\mu(q)$ is nonlinear \begin{equation}
\nu(q)\equiv\frac{\mu(q)}{q}\ne const\:,\label{eq:strong}\end{equation}
 transport is multifractal. This distinction is important since in
the weak case the PDF evolves in a self-similar way: \begin{equation}
P(s,t)=t^{-\nu}f(\xi),\hspace{1cm}\xi\equiv t^{-\nu}(s-\langle s\rangle)\label{selfsim}\end{equation}
 while a non-constant $\nu(q)$ in (\ref{eq:strong}) precludes such
self-similarity (see the discussions in \cite{Kuznetsov2000,LKZ01}
for details about the non self-similar behavior). 

In order to characterize transport, we focus on the arc length $s(t)$
of the path traveled by an individual tracer up to a time $t$,\begin{equation}
s_{i}(t)=\int_{0}^{t}v_{i}(t')dt'\:,\label{arclenghtdefinition}\end{equation}
where $v_{i}(t')$ is the absolute speed of the particle $i$ at time
$t'$. This choice is motivated by the fact that in order to consider
mixing properties from the dynamical principles it is important to
consider the trajectories within the phase space. However, for the
system of passive particles the phase space is formally {}``identical''
to the physical space where particles evolve (see Eq.\ref{eq:Tracers_Hamiltonian}).
Another important feature of the arc length is that it is independent
of the coordinate system and as such we can expect to infer intrinsic
properties of the dynamics. Moreover, expression (\ref{arclenghtdefinition})
implies that $s_{i}(t)$ reflects the history up to time $t$ of the
speed $v_{i}$. Since the velocities are bounded the observations
described further on of anomalous transport behavior are directly
linked to strong memory effects.

\subsection{Particle Transport}

\begin{figure}
\includegraphics[%
  width=7cm,
  keepaspectratio]{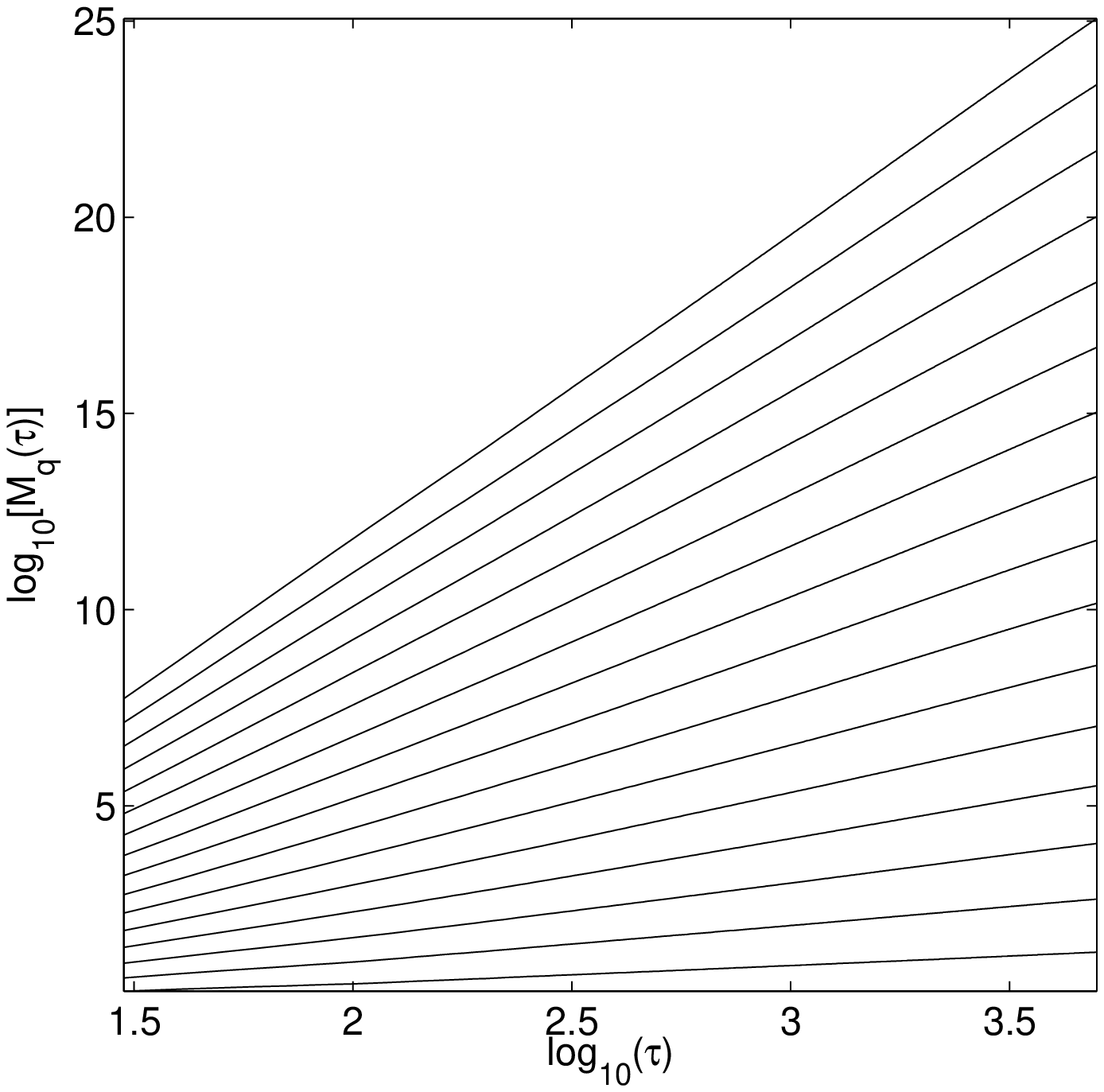} \\
\includegraphics[%
  width=7cm,
  keepaspectratio]{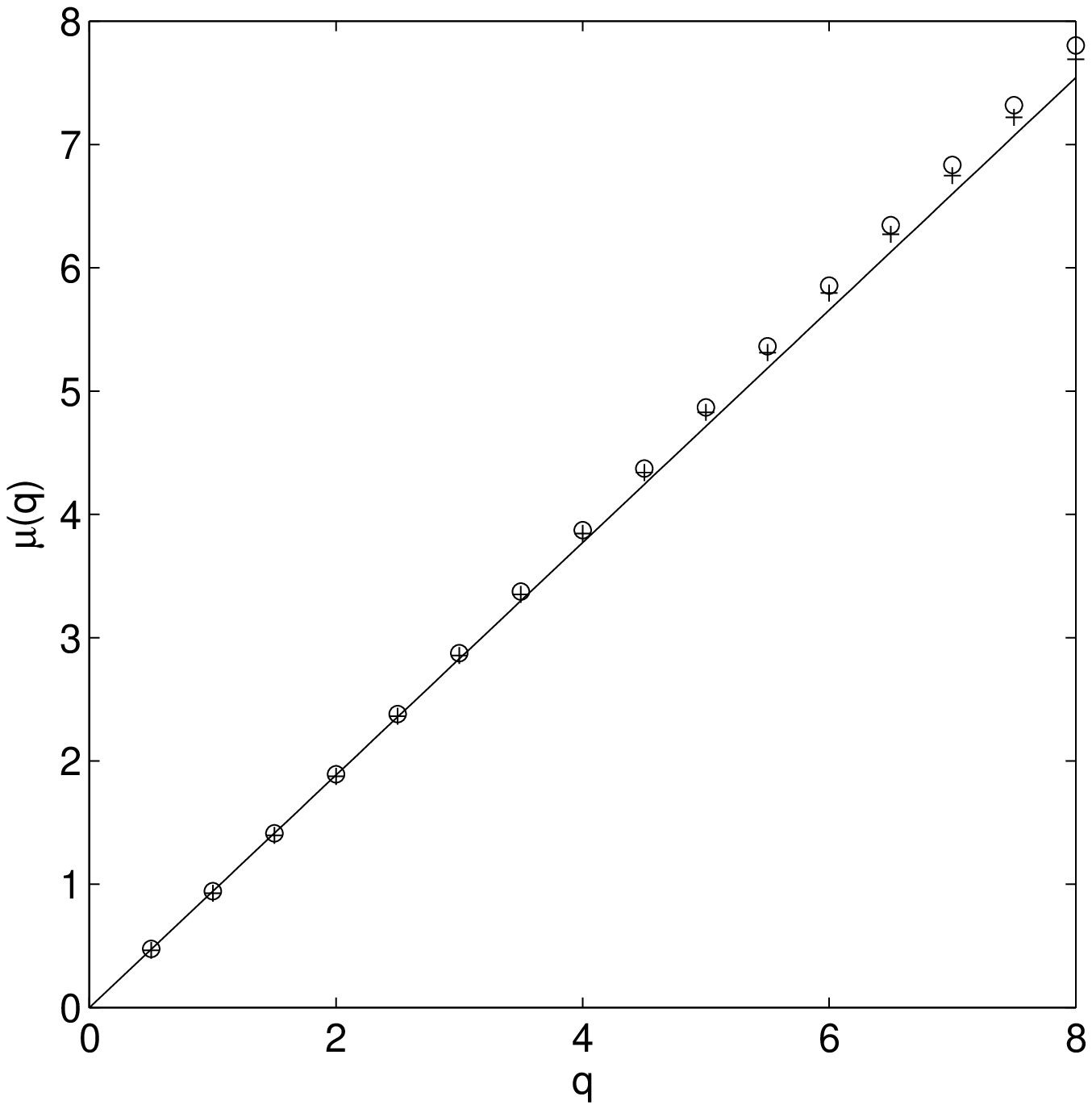}

\caption{\label{cap:Transport_1}Top: Moments of distribution of the arc length
$M_{q}(\tau)=\langle|s(\tau)-\langle s(\tau)\rangle|^{q}\rangle$
versus time of tracers evolving in the smooth field for $q=1/2,\:1,\:3/2,\cdots8$.
The behavior $M_{q}(\tau)\sim\tau^{\mu(q)}$ is confirmed. Bottom:
Characteristic exponent versus moment order, $q$ vs $\mu(q)$, with
$\mu(2)=1.87$. o and + symbols refer respectively to computations
made using 512 and 256 particles. The solid line corresponds to $\mu(q)=\mu(1)q$
expected for self-similar behavior. Transport is super-diffusive and
multifractal.}
\end{figure}
\begin{figure}
\includegraphics[%
  width=7cm,
  keepaspectratio]{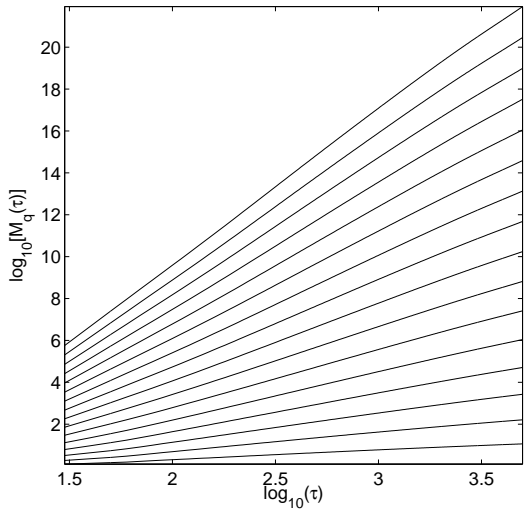}\\
\includegraphics[%
  width=7cm,
  keepaspectratio]{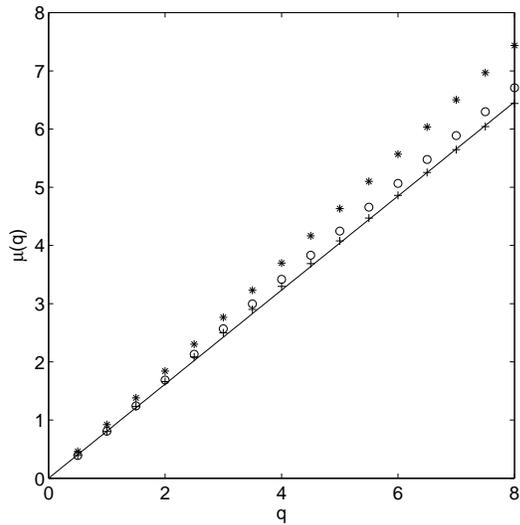}

\caption{\label{cap:Transport_2}Top: Moments of distribution of the arc length
$M_{q}(\tau)=\langle|s(\tau)-\langle s(\tau)\rangle|^{q}\rangle$
versus time of tracers evolving in the forced field for $q=1/2,\:1,\:3/2,\cdots8$.
The behavior $M_{q}(\tau)\sim\tau^{\mu(q)}$ is confirmed. Bottom:
Characteristic exponent versus moment order, $q$ vs $\mu(q)$, with
$\mu(2)=1.68$. One can notice a change of slope after $\log_{10}(\tau_{0})=3.3$.
The o and + symbols refer respectively to computations made using
512 and 256 particles for $\tau>\tau_{0}$, while the {*} sign corresponds
to $\tau<\tau_{0}$. The solid line corresponds to $\mu(q)=\mu(1)q$
expected for self-similar behavior. Transport is super-diffusive.}
\end{figure}
\begin{figure}
\includegraphics[%
  width=7cm,
  keepaspectratio]{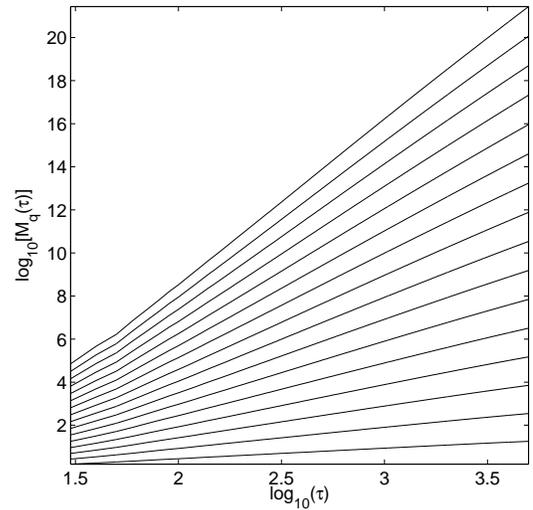}\\
\includegraphics[%
  width=7cm,
  keepaspectratio]{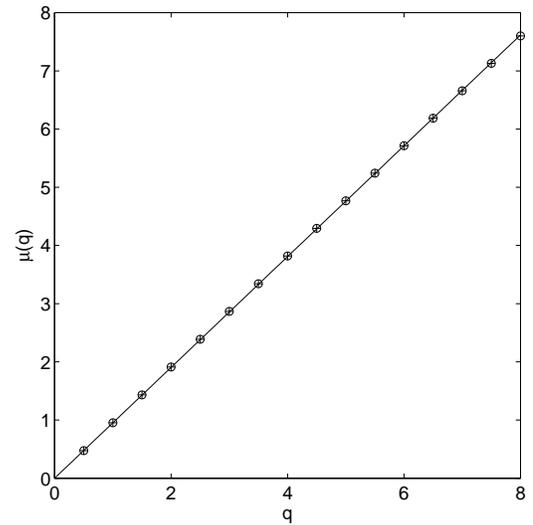}

\caption{\label{cap:Transport_3}Top: Moments of distribution of the arc length
$M_{q}(\tau)=\langle|s(\tau)-\langle s(\tau)\rangle|^{q}\rangle$
versus time of tracers evolving in the anisotropic field for $q=1/2,\:1,\:3/2,\cdots8$.
The behavior $M_{q}(\tau)\sim\tau^{\mu(q)}$ is confirmed. Bottom:
Characteristic exponent versus moment order, $q$ vs $\mu(q)$, with
$\mu(2)=1.85$. o and + symbols refer respectively to computations
made using 512 and 256 particles. The solid line corresponds to $\mu(q)=\mu(1)q$
expected for self-similar behavior. Transport is super-diffusive and
single fractal.}
\end{figure}
We followed the evolution of $512$ passive particles as defined by
Eq.(\ref{gen.adv}), and we computed $s_{i}(t)$'s up to a time $\tau=10^{4}$.
In order to keep a constant numerical accuracy the increments $\Delta s_{i}(nT)=s_{i}((n+1)T)-s_{i}(nT)$
were recorded for the successive diagnostic times $nT$. This last
feature has also the advantage of providing better statistics when
computing moments. Indeed since we have chosen stationary regimes
for the field, we can assume that transport properties are independent
of the initial condition of the field and thus use time-invariance
to increase statistics. This time invariance becomes however not useful
for times close to $\tau=10^{4}$, and statistics in this region is
thus not accurate anymore, also we are not interested in short time
behavior. We have therefore considered times $10^{2}<\tau<5.10^{3}$
in order to compute the $\mu(q)$.

In all three considered cases the transport is found to be anomalous
and superdiffusive, with a characteristic second order exponent $\mu(2)\approx1.8$
for all cases. The time behavior of the moments and characteristic
exponents for all considered cases are plotted in figures \ref{cap:Transport_1},
\ref{cap:Transport_2} and \ref{cap:Transport_3}, and a summary is
provided in table \ref{cap:table-2}, where we included for comparison
the results obtained in Ref. \cite{LZ02} for a flow governed by point
vortices. The similarity observed for the exponents in these quite
different settings of the parameters and regimes of the Charney-Hasegawa-Mima
equation as well as the one observed in point vortices, points to
a universal behavior for the transport of passive tracers in these
two-dimensional flows, which in some sense confirms the validity of
the quite general estimation of $\mu\approx1.5$ proposed in \cite{LKZ01}
as a value for an universal exponent. 

However, conversely to the point vortices case, one can notice from
Figs. \ref{cap:Transport_1}, \ref{cap:Transport_2} and \ref{cap:Transport_3}
that the behavior of characteristic exponent versus moment for the
first two cases may correspond more to the multifractal type although
the nonlinearity is quite weak, and error bars are quite large. For
the third case, transport is more or less simply fractal. This behavior
corresponds to simple fractal transport in the anisotropic system,%
\begin{table}

\caption{\label{cap:table-2}Characteristic second moment exponent for the
three different cases studied. Exponents obtained in flows governed
by point vortices in Ref. \cite{LZ02} are given for comparison. }

\begin{tabular}{|c|c|}
\hline 
Point vortices &
\begin{tabular}{cc}
4 vortices&
$\mu(2)\approx1.82$\tabularnewline
16 vortices &
$\mu(2)\approx1.77$\tabularnewline
\end{tabular}\tabularnewline
\hline 
Charney-Hasegawa-Mima&
\begin{tabular}{cc}
Smooth Field&
$\mu(2)\approx1.81$\tabularnewline
Forced Field&
$\mu(2)\approx1.73$\tabularnewline
Anisotropic Field&
$\mu(2)\approx1.85$\tabularnewline
\end{tabular}\tabularnewline
\hline
\end{tabular}
\end{table}
 which implies an almost self-similar behavior of the distribution
function. In fact as may be illustrated in figure \ref{cap:Transport_2},
we may expect that with more particles and larger times the multifractal
behavior may be more clear, a feature which can also be the case for
the third case. When computing characteristic exponents one also has
to recall the possibility of log-periodic oscillations \cite{Benkadda99}.
This phenomenon may also be responsible for the uncertainties on the
measured values of the exponents. 

All in all these results imply that at least for intermediate times,
transport is anomalous and single fractal, which means that transport
properties in these systems should be correctly described by a fractional
Fokker-Plank-Kolmogorov equation of the type \cite{Zaslavsky2002}:\begin{equation}
\frac{\partial^{\beta}P(s,t)}{\partial t^{\beta}}=\mathcal{D}\frac{\partial^{\alpha}P(s,t)}{\partial|s|^{\alpha}}\:,\label{Eq:Fractional}\end{equation}
with $\mu\approx2\beta/\alpha$ (see for instance \cite{LKZ01,Zaslavsky2002}).
For larger time transport may remain single fractal or develop a multifractal
behavior in which case a model of transport properties using Eq.(\ref{Eq:Fractional})
may be only approximate. Note that using Eq.(\ref{Eq:Fractional})
to model transport properties implies that a kinetic limit has been
performed on particle statistics. In this situation we are dealing
with Levy type processes, hence moments higher than two are likely
not defined. This particularity is however linked to the kinetic limit,
and since we are dealing numerically with a finite number of particles
during a finite time, and also since velocities are bounded, moments
of particle arc-length distribution are defined, see \cite{Zalavsky2000}
for a discussion on the matter.

\section{Jets\label{sec:Jets}}

\subsection{Definitions}

Tracking the origin of anomalous transport can be fairly well understood
when one is able to draw a phase portrait using a Poincare map and
for instance to measure Poincare return time to a given region of
phase space. The conclusion of this type of analysis will almost certainly
lead to the fact that the phenomenon of stickiness on the boundaries
of the islands generates strong {}``memory effects'', as a result
of which transport becomes anomalous. However, when dealing with a
more complex system, for which the drawing of a phase portrait is
not achievable, one has to rely on other techniques.%
\begin{figure}
\begin{center}\includegraphics[%
  width=7cm]{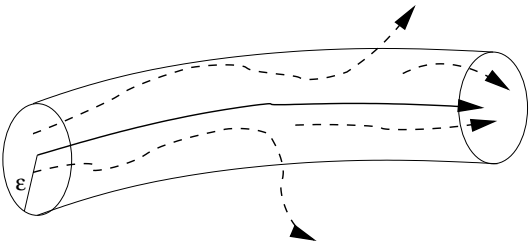}\end{center}

\caption{Tracking of $\epsilon$ coarse-grained regular jet.\label{cap:Tracking-of-jet}}
\end{figure}

In a two dimensional phase-space, the phenomenon of stickiness corresponds
often to passive particles remaining for large times in the neighborhood
of an island of regular motion. A consequence of this behavior is
that sticky zones are regions where particles are trapped, and therefore
are regions where particles remain in each others neighborhood for
large times. It becomes then natural when dealing with more complex
systems for which no phase portrait can easily be drawn, to look for
places where passive particles remain in each other's vicinity for
large times. One possibility to capture this feature of the dynamics
is to look for its signature by measuring finite time Lyapunov exponents
(FTLE), and by isolating within the space of initial conditions, regions
of vanishing FTLE's (see for instance \cite{Boatto99}). This type
of approach has however its shortcoming, namely sticky regions are
not necessarily smooth from the microscopic point of view, meaning
they can be regions of strong chaos that are somehow restricted within
an arbitrary small scale, which may be problematic when dealing with
FTLE's. The other possibility is to look directly for chaotic jets
\cite{LZ02}. These chaotic jets can be understood as moving clusters
of particles within a specific domain for which the motion appears
as almost regular from a coarse grained perspective. From another
point of view, looking for chaotic jets can be understood as a particular
case of measurements of space-time complexity\cite{Afraimovich03}.

In order to look for jets we proceed as described in the illustration
presented in Fig. \ref{cap:Tracking-of-jet} which provides an easy
and intuitive description of the mechanism used to detect jets. To
summarize, we consider a reference trajectory $\mathbf{r}(t)$ within
the phase space. We then associate to this trajectory a corresponding
{}``coarse grained'' equivalent, i.e the union of the balls $\cup B(\mathbf{r}(t),\epsilon)$
of radius $\epsilon$ whose center is the position $\mathbf{r}(t)$.
Given an $\epsilon$-coarse grained trajectory, we analyze the behavior
of real trajectories starting from within the ball at a given time,
and measure the time $\tau$ and length $s$, before the trajectory
escapes from the coarse grained trajectory. We then analyze the resulting
distributions. This approach has already been used with success when
studying numerically the advection of passive tracers in flows governed
by point vortices \cite{LZ02}. The main difficulty in using this
diagnostic follows from the fact that data acquisition is not sampled
linearly in time nor space, a point which leads in the present case
to some difficulties.

\subsection{Statistical results}

In the setting of the evolution of passive tracers within the three
considered Charney-Hasegawa-Mima flows, we settled for the following
values $\epsilon=10^{-1}$ and $\delta=10^{-3}$. These values are
to be compared with the $\epsilon=10^{-2}$ and $\delta=10^{-6}$
considered for point vortex systems in \cite{LZ02}. %
\begin{figure}
\includegraphics[%
  width=7cm,
  keepaspectratio]{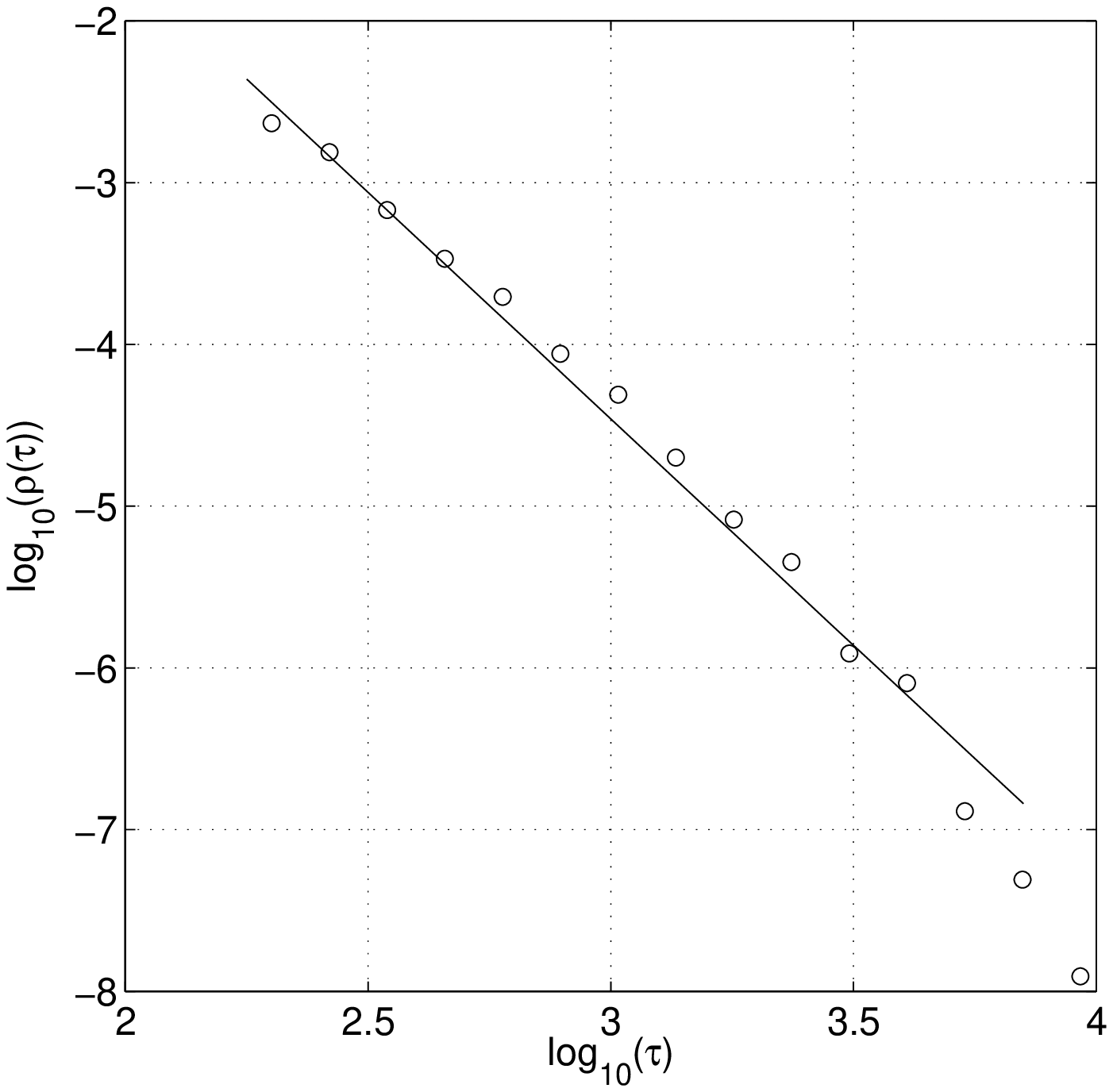}

\caption{\label{cap:Trapping_times_1}Tail of the distribution of trapping
times $\tau$ for the smooth field (Fig.~\ref{cap:Case1}). A power-law
decay is observed with typical exponent $\rho(\tau)\sim\tau^{-\gamma}$
with $\gamma\approx2.8\pm0.1$. }
\end{figure}
\begin{figure}
\includegraphics[%
  width=7cm,
  keepaspectratio]{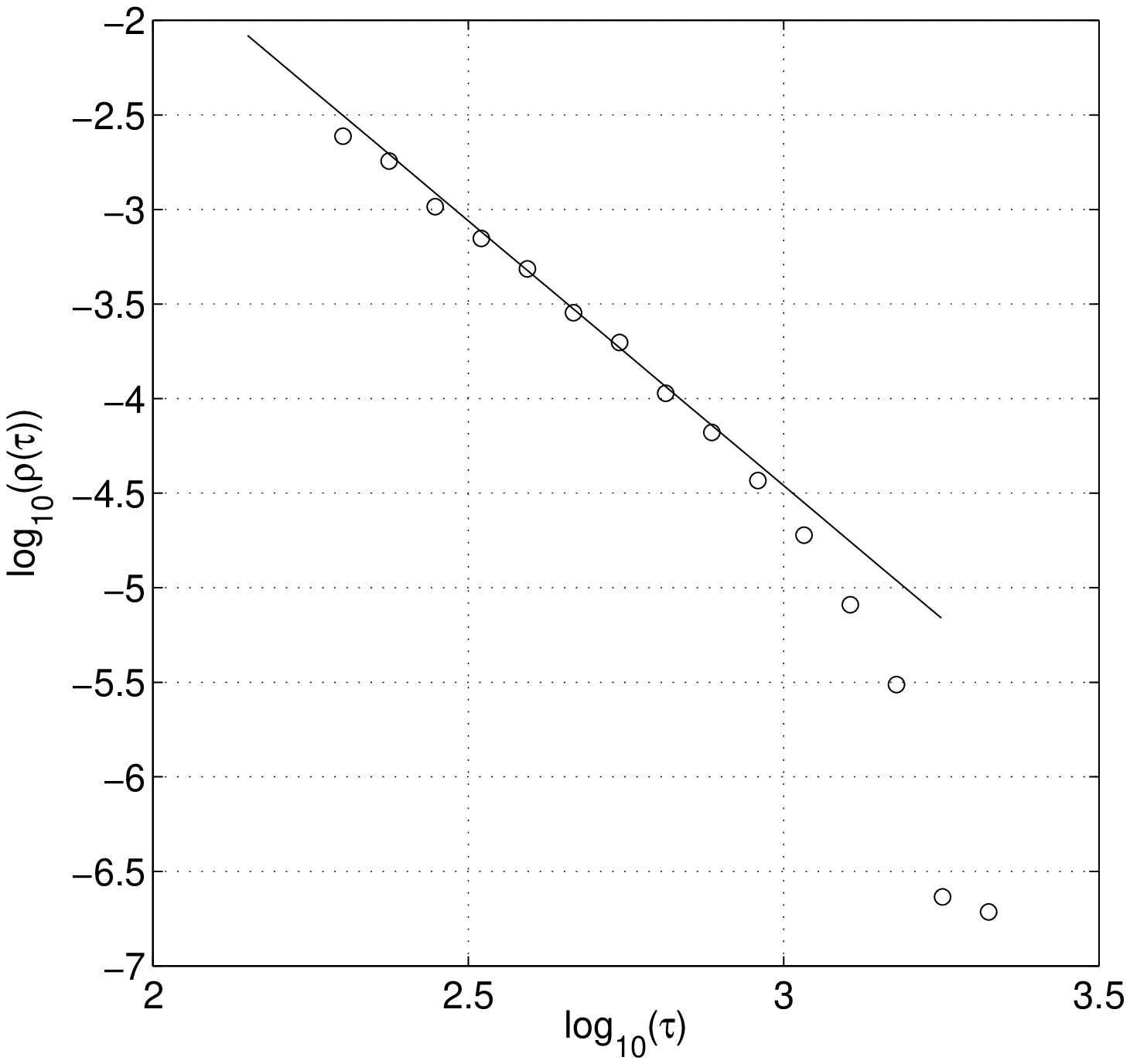}

\caption{\label{cap:Trapping_times2}Tail of the distribution of trapping
times $\tau$ for the forced field (Fig.~\ref{cap: Case 2}). A power-law
decay is observed with typical exponent $\rho(\tau)\sim\tau^{-\gamma}$
with $\gamma\approx2.8\pm0.1$. }
\end{figure}
\begin{figure}
\includegraphics[%
  width=9cm,
  keepaspectratio]{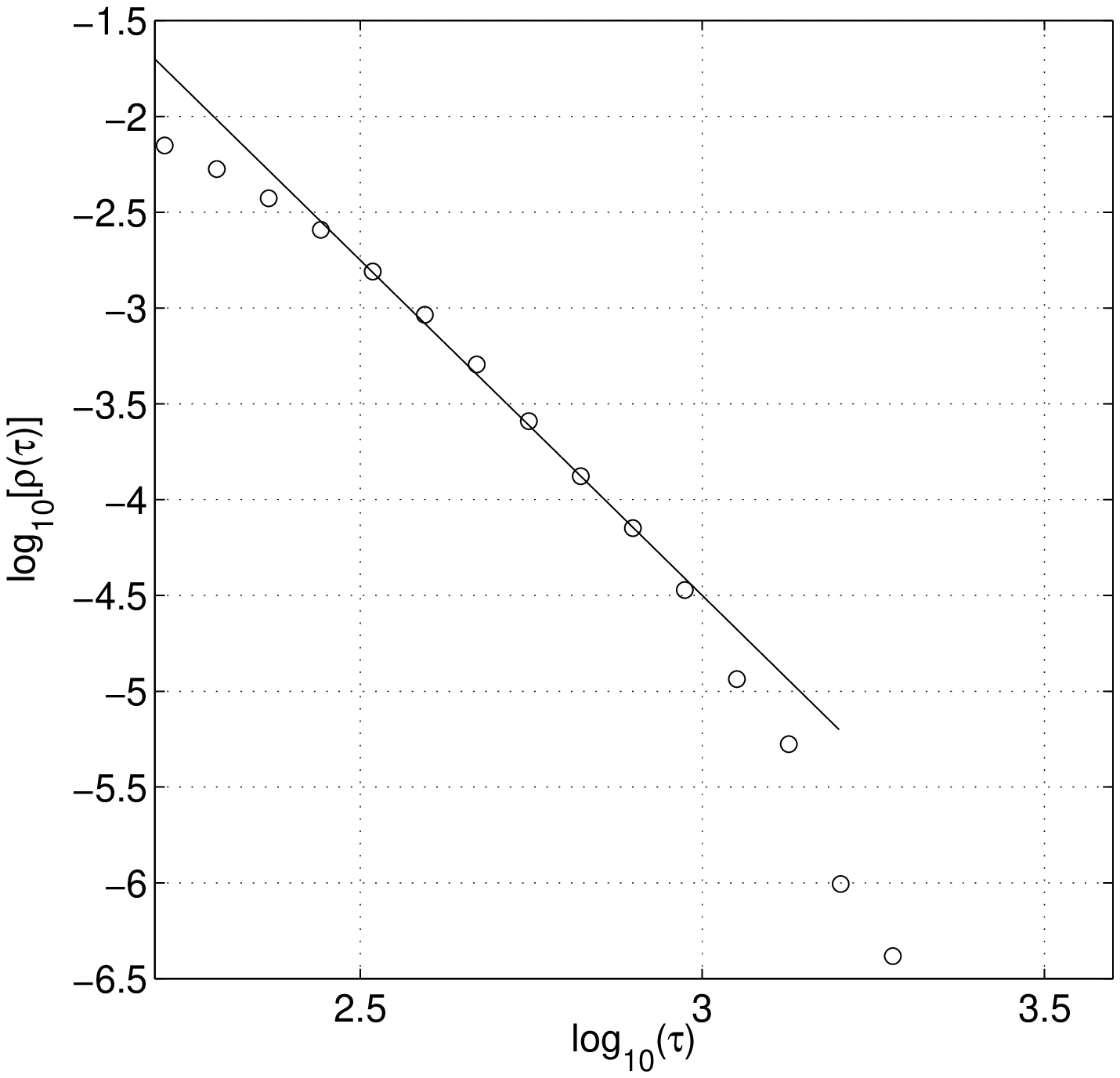}

\caption{\label{cap:Trapping_times3}Tail of the distribution of trapping
times $\tau$ for the anisotropic field (Fig.~\ref{cap: Case 3}).
A power-law decay is observed with typical exponent $\rho(\tau)\sim\tau^{-\gamma}$
with $\gamma\approx3.5\pm0.5$. }
\end{figure}
 This reduction of accessible scales stems from the fact that for
the length of our simulations ($\tau=10^{4}$), only a few trajectories
are escaping from the jets when we try to use the point vortex values.
This leaves us with not enough data to gather realistic statistics.
It is important to note that results should not depend much on the
value of $\epsilon$ (see the discussion in \cite{LZ02}), as long
as its value does not cross a characteristic scale in the system.
For instance in the point vortex systems, cores surrounding vortices
had a radius of order $\sim0.1$, or here in the forced systems, the
characteristic wavelength corresponds to scales of order $\sim1$.
Moreover in order to infer some possible fractal structure of the
jet, we are constrained to consider the largest possible $\epsilon$
once $\delta$ has been fixed. Here the choice made for $\delta$
was the smallest we could make in order to gather enough statistics,
although these may not be sufficient for the anisotropic case. We
also tried to keep at least two orders of magnitude between $\epsilon$
and $\delta$, to infer eventual fractal properties.

With the present choice of parameters we were able for gather about
$15000$ events for each of the three cases. We followed $256$ reference
trajectories for a time $\tau=10^{4}$ during which the behavior of
two nearby tracers was checked. We call a reference trajectory the
trajectory of a given passive tracer which evolves freely, we then
put randomly at a distance $\delta$ of this tracer a second (or more)
tracer which was called a ghost in \cite{LZ02}. We then let these
tracer evolve until the distance $\epsilon$ is reached. Then the
ghost is removed and time interval $\Delta\tau$ and travelled length
$\Delta s$ are recorded. A new ghost is assigned to the reference
trajectory and so on. We of course expect that the portion we compute
of reference trajectory ($\tau=10^{4}$) is sufficiently ergodic in
the accessible phase space.

We were then able to obtain the trapping time distributions $\rho(\tau)$
described in Figs.~\ref{cap:Trapping_times_1}, \ref{cap:Trapping_times2}
and \ref{cap:Trapping_times3}. The characteristic exponent $\rho(\tau)\sim\tau^{-\gamma}$
for trapping times observed in two out of the three systems is typically
$\gamma\approx2.8$. For these cases we therefore obtain good agreement
with the \begin{equation}
\gamma\approx\mu(2)+1\:,\label{eq:gamma_mu}\end{equation}
 relation which links the transport exponent $\mu$ to the characteristic
trapping time exponent. The law (\ref{eq:gamma_mu}) can be linked
to the fractional transport equation (\ref{Eq:Fractional}). One can
link the $\gamma$ exponent to the $\beta$ exponent as follows $\beta=\gamma-1$
, and $\mu=2\beta/\alpha$ (see \cite{LKZ01,LZ02} and references
therein). $\alpha$ being linked to the spatial fractal properties,
the equation (\ref{eq:gamma_mu}) is valid when $\alpha\sim2$. One
can also notice in Fig.s.~\ref{cap:Trapping_times_1}, \ref{cap:Trapping_times2}
that the law $\rho(\tau)\sim\tau^{-\gamma}$ applies typically for
times up to $\tau\sim1000$ or a little more. This behavior can be
expected as we are following a finite number of reference trajectories
($256$) during a finite time ($\tau_{max}=10^{4}$). In order to
potentially obtain a broader range of applicability one, would have
to increase the value of $\tau_{max}$ to at least $10^{5}$. This
is unfortunately beyond our computing resources. Note also, that due
to the inverse cascade there is an accumulation of energy at large
scales, such that the driving field may not always be considered to
have reached a stationary state if we decided to increase simulation
length. In order get better statistics, one also has to be careful
in not adding to many particles instead of increasing time. Indeed
the limits $N\rightarrow\infty$ and $\tau\rightarrow\infty$ are
most likely not commuting in these anomalous regimes where rare events
related to memory effects are important.

In fact the law (\ref{eq:gamma_mu}) also applies in the anisotropic
field for small trapping times, but o. verth whome range it is clear
in Fig.~\ref{cap:Trapping_times3} that no scaling law really emerge.
When comparing with the two other cases, we may find two possible
reason for this failure. First, we can notice that in this configuration
of the field the energy and enstrophy are quite low, implying weak
nonlinear effects. It may thus take more time to fill the tail of
the trapping time distribution. Second, another peculiarity of the
anisotropic case is the the fact pointed out in \cite{Basu03,Basu_03}:
in the anisotropic case, the conservation of the generalized vorticity
$\Omega=\Phi-\lambda\Delta\Phi+gx\:,$ by the evolution of the Hasegawa-Mima
equation (\ref{eq:H-M}) implies necessarily that for strong values
of $g$ the motion along the $x-$direction is bounded. Hence in this
situation the motion of passive tracers is quasi one dimensional.
In this setting it is likely that the hidden fractal properties in
the system may differ than for a two-dimensional systems, especially
regarding the derivation of the laws linking $\beta$, $\gamma$ and
$\alpha$. In these regards, the fact that the law (\ref{eq:gamma_mu})
is valid for small times up to a time $\tau_{c}$ may just be a simple
consequence of the fact that for $\tau<\tau_{c}$, the passive tracers
have not yet reached the boundaries imposed by $g$ and a spatially
two dimensional behavior is still accurate. At this point it is important
to recall that we have periodic boundary conditions. The constraint
imposed by the conservation of $\Omega$ is thus not as important
for small values of $g$ corresponding to the first two cases.

\begin{figure}
\includegraphics[%
  width=7cm,
  keepaspectratio]{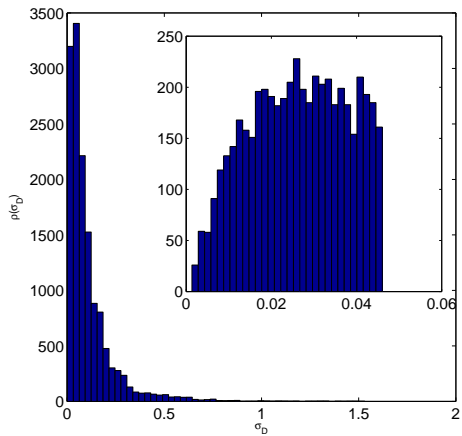}

\caption{\label{cap:Lyapun_dist} Unnormalized Distribution of Lyapunov exponents
$\sigma_{D}$, see Eq. (\ref{lyapunivarclength}). An accumulation
towards zero is observed. The zoom near small regions reveals an actual
decrease for $\sigma_{D}<0.02$, such behavior is expected as data
is bounded by finite speed and finite time.}
\end{figure}

The rarity of events with smaller values of $\delta$ indicates that
the trajectories of tracers are relatively regular when one considers
small scales, which implies the possibility of vanishing Lyapunov
exponents. Hence, following the definitions given in \cite{LZ02}:
\begin{eqnarray}
\sigma_{L} & = & \frac{1}{\Delta t}\ln\frac{\epsilon}{\delta}\:,\label{lyapunovtime}\\
\sigma_{D} & = & \frac{1}{\Delta s}\ln\frac{\epsilon}{\delta}\:,\label{lyapunivarclength}\end{eqnarray}
we compute the distribution of Lyapunov exponents $\sigma_{D}$ obtained
from the smooth case data in Fig.~\ref{cap:Lyapun_dist}%
, where one can directly see an accumulation towards zero of the distribution.
The zoom in Fig.~\ref{cap:Lyapun_dist} shows an actual decrease
for $\sigma_{D}<0.02$. This behavior is expected as we only have
finite speeds in the system and simulations are carried for a finite
time. For instance $\sigma_{D}<0.02$ corresponds to an average trapping
time $t_{a}=1100$. This accumulation of exponents toward zero is
quite important as it is a strong evidence of memory effects. The
system does not display universal hyperbolic properties, and thus
transport modeling using this as a hypothesis are probably doomed.
This also shows the relevance of the notion of weak chaos and weak
mixing properties, and therefore the necessity of considering simple
models such as the billiard considered in \cite{Zaslav2001}. Indeed
these provide good insights on how the presence of quasi zero Lyapunov
exponents does not necessary mean the system is regular, but that
complexity is not growing as fast as expected \cite{Afraimovich03}.

\subsection{Localization and structure of jets}

\begin{figure}
\includegraphics[%
  width=7cm,
  keepaspectratio]{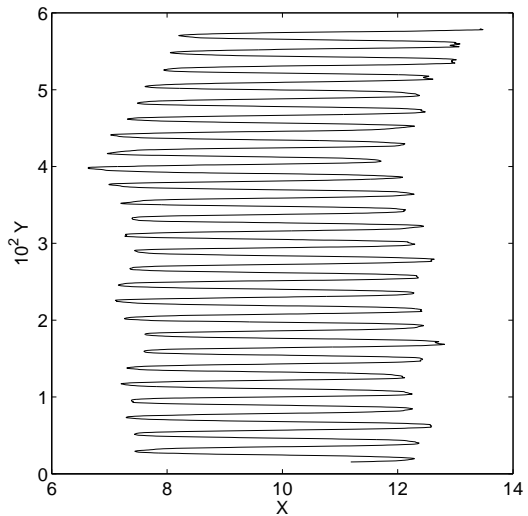}

\caption{\label{cap:jet_Location} Localization of a long lived jet in the
{}``Forced Field'' (cf Fig. \ref{cap: Case 2}). The jet is bouncing
back and forth between the two perturbed vortices}
\end{figure}
Due to the shape of the finite size Lyapunov exponent distributions
observed, it is also possible to track and localize jets which are
responsible for the anomalous behavior. Indeed, these exponents are
monotonically decreasing. Hence we can define a threshold beyond which
the jet can be considered {}``regular''. We can track the reference
trajectory afterwards in order to localize regions responsible for
anomalous behavior. In Fig.~\ref{cap:jet_Location} we show the localization
of a jet for which the trapping time of nearby tracers is found to
be $\approx1000$, in the case of the forced field (cf Fig.~\ref{cap: Case 2}).
The jet is bouncing back and forth between the two perturbed vortices.
\begin{figure}
\includegraphics[%
  width=7cm,
  keepaspectratio]{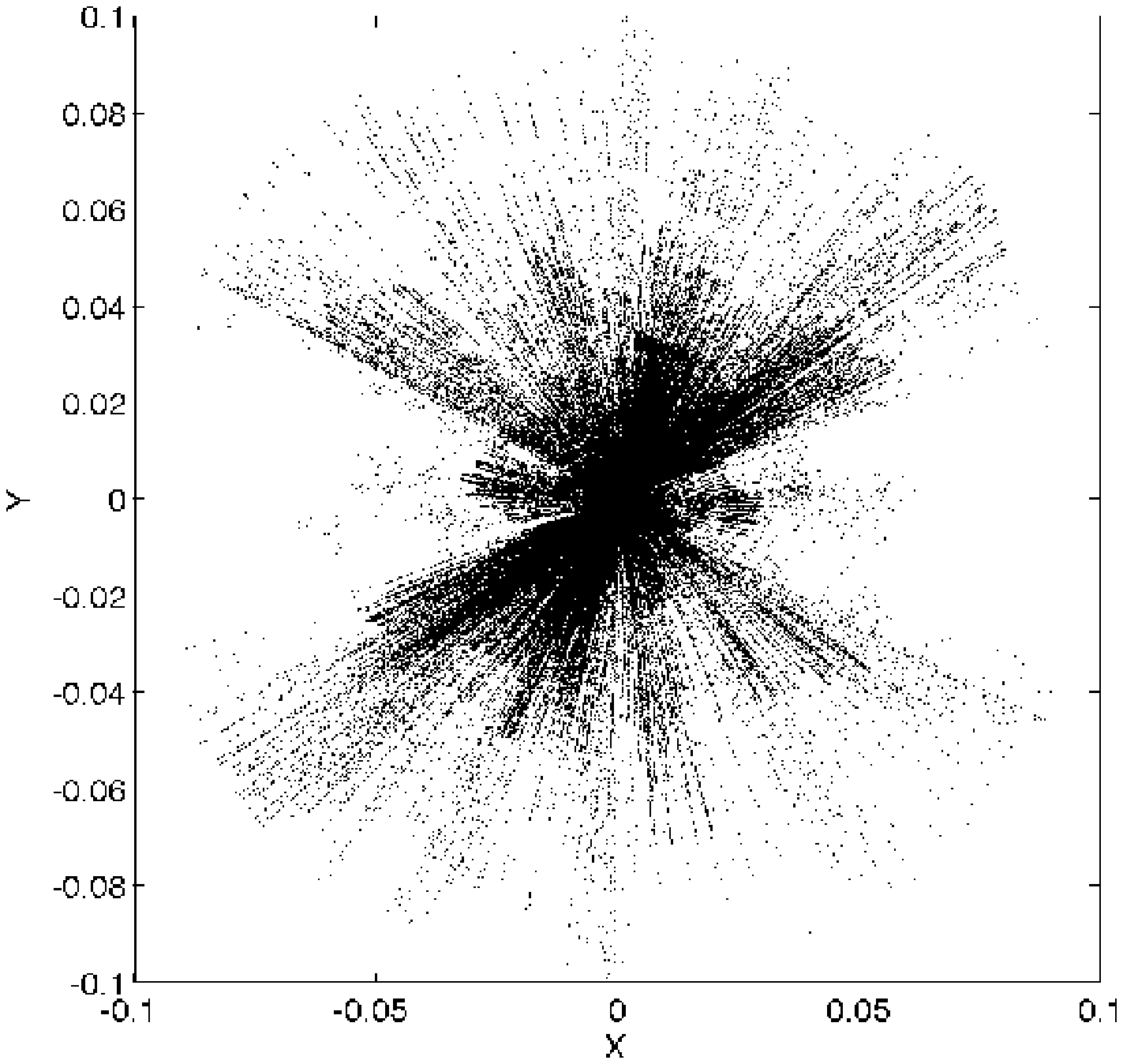}\\
\includegraphics[%
  width=7cm]{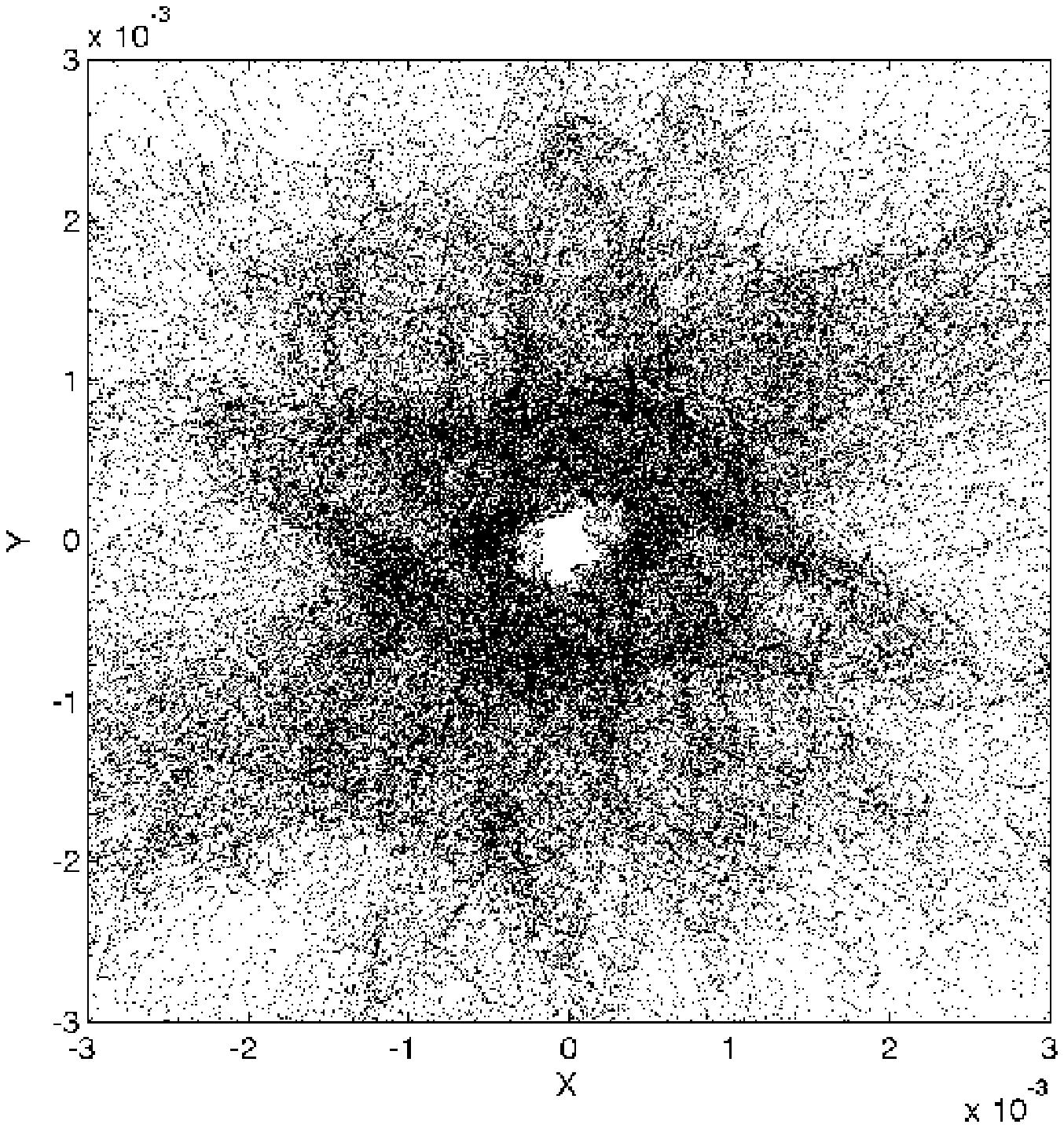}

\caption{\label{cap:jet_struct} Relative position of test particles for the
long lived jet depicted in Fig. \ref{cap:jet_Location} (top). Zoom
(bottom), test particles reach distances $<10^{-4}$ to the reference
trajectory.}
\end{figure}
The procedure to detect a jet is quite simple, we chose a reference
trajectory and start to compute trapping times of ghosts. When we
notice that a ghost has not escaped for a while (we choose $\tau>200$
for this jet), we start recording the positions of the reference tracer.
We also reset the ghosts and add many more in order to track the {}``structure''
of the jet, and record their positions as well. When one of the ghosts
leave the jet, we measure the actual trapping time. If it appears
long enough (here we chose $\tau>1000$), we stop, else we reset everything
and wait until the reference trajectories detect a new potential jet.
The advantage of this method is that we have no a priori information
on the jet location. For instance the jet depicted in Fig.~ \ref{cap:jet_Location},
was not anticipated, as we expected more a trapping within a vortex
or in its neighborhood. We may notice however that by using this method
of detection we miss the first portion of the jet. Note also that
by localizing jets, we are able to detect regions responsible for
anomalous behavior of transport properties. It therefore may therefore
give a good clue on where and how to act on a system, if one wants
to reduce or eliminate this anomaly.

Once a jet is located, one can infer the behavior of the test tracers
with respect to the reference trajectory. In order to get Fig.~\ref{cap:jet_struct},
256 test particles were initialized on a circle around a reference
particle whose trajectory was trapped in the long lived jet localized
in Fig.~\ref{cap:jet_Location}, and the relative positions of the
test particles were plotted during the life of the jet. One notices
a {}``star'' like shape of the figure, this phenomenon is a consequence
of multiple stretching and squeezing of the circle of particles, which
one can observe while looking dynamically at the evolution of the
circle. One also can notice that ghosts which initially are localized
at a distance $\delta=10^{-3}$ from the reference tracer, can find
themselves much closer to the test particles close to $10^{-4}$ in
Fig.~\ref{cap:jet_struct}. This implies that we do not expect that
the value of the trapping time exponent $\gamma$ is strongly dependent
on the initial choice of $\delta$. Note also that the behavior portrayed
in Fig.~\ref{cap:jet_struct} seems to indicate that chaos is present
at these scales, thus contrary to our physical intuition the relative
motion within the jet is not regular. However despite this fact, the
chaotic behavior of trajectories is trapped within a given small scale
for a long time, which in the end gives rise to a regular non-diffusive
trajectory from a coarse grained perspective. This trapping phenomena
is also illustrated by the apparently {}``quantified'' maximum radii
seen on the star shape displayed in Fig.~\ref{cap:jet_struct}. We
may thus speculate that these radii correspond to actual successive
barriers which are restricting chaos within a small scale. Note also
that the chaotic behavior within the jet may have induced numerical
computations of strong values for finite time Lyapunov exponents.
In this situation relying on these exponents would have meant that
jets which are responsible for anomalous transport may go unnoticed.

\section{Conclusion\label{sec:Conclusion}}

In this paper we have investigated the dynamical and statistical properties
of passive particle advection in different configurations of Charney-Hasegawa-Mima
flows. The goal of the work was to consider transport properties of
these systems while putting in perspective the results obtained for
point vortex flows. In this sense it was a step further in providing
qualitative insights on general transport properties of two-dimensional
flows. The transport properties of all the considered cases are found
to be anomalous with characteristic exponent $\mu\sim1.75-1.8$. These
values are also quantitatively comparable to the results obtained
for point vortex flows. 

In order to analyze the origin of the anomalous transport properties,
passive tracer motion is analyzed by measuring the mutual relative
evolution of two nearby tracers, i.e by looking for chaotic jets \cite{LZ02}.
The jets can be understood as moving clusters of particles within
a specific domain where the motion is almost regular from a coarse
grained perspective, inducing memory effects and long time-correlations.
The distribution of trapping times in the jets shows a power-law tail
whose characteristic exponent is in very good agreement with the law
$\gamma=\mu+1$ linking the transport exponent $\mu$ to the trapping
time exponent $\gamma$. This agreement is a good signature that the
origin of anomalous transport in these system is intimately related
to the existence of jets in the sense described previously. The localization
of jets in the system can be done, and it is shown that jets are not
necessarily located around a coherent structure as was the case for
point vortex flows, but that they can manifest themselves by a trajectory
bouncing back and forth between 2 structures. Moreover when analyzing
the {}``structure'' of the jet, it is shown that the trajectory
of individual tracers are likely chaotic within the jet, but that
this chaotic behavior is for long times restricted within a given
small scale, giving rise to the regular non-diffusive structure.

\begin{acknowledgments}
We would like to thank D. F. Escande for discussions, as well as corrections
and suggestions regarding the manuscript. Part of the work presented
in this paper was carried out  while X.L was visiting the Department
of Fundamental Energy Science, Graduate School of Energy Science,
Kyoto University. X.L thanks S. Hamaguchi and A. Bierwage for useful
discussions as well as the Department for financial support. G.M.Z.
was supported by the U.S. Navy Grant N00014-96-1-0055 and the  U.S.
Department of Energy Grant No. DE-FG02-92ER54184.
\end{acknowledgments}
\bibliographystyle{unsrt}
\bibliography{./transport}

\end{document}